\documentclass[12pt,preprint]{emulateapj}

\usepackage{color}
\usepackage{graphicx}
\usepackage{natbib, aas_macros}
\bibpunct[,]{(}{)}{;}{a}{}{,}
 \usepackage{multirow}
\usepackage{amsmath,amssymb,mathrsfs}
\shortauthors{HAYASHI \& CHIBA.}
\shorttitle{A constant surface density in dSphs}

\begin{document}

\title{A common surface-density scale for the Milky Way and Andromeda dwarf satellites as a constraint on dark matter models}

\author{Kohei~Hayashi\altaffilmark{1} and Masashi~Chiba\altaffilmark{1}}

\altaffiltext{1}{Astronomical Institute, Tohoku University,
Aoba-ku, Sendai 980-8578, Japan \\E-mail: {\it k.hayasi@astr.tohoku.ac.jp; chiba@astr.tohoku.ac.jp}}
\begin{abstract}
In an attempt to place an explicit constraint on dark matter models, we define and estimate the mean surface density of a dark halo within the radius of the maximum circular velocity, which is derivable for various galaxies with any dark matter density profiles.
We find that this surface density is generally constant across a wide range of maximum circular velocities of $\sim$ 10--400 km~s$^{-1}$, irrespective of the different density distributions in each of the galaxies. 
This common surface density at high halo-mass scales is found to be naturally reproduced by both cold and warm dark matter (CDM and WDM, respectively) models, even without employing any fitting procedures.
However, the common surface density at dwarf-galaxy scales, for which we have derived from the Milky Way and Andromeda dwarf satellites, is reproduced only in a massive range of WDM particle masses, whereas CDM provides a reasonable agreement with the observed constancy.
This is due to the striking difference between mass--concentration relations for CDM and WDM halos at low halo-mass scales.
In order to explain the universal surface density of dwarf-galaxy scales in WDM models, we suggest that WDM particles need to be heavier than 3~keV.
\end{abstract}


\section{INTRODUCTION}

Cold dark matter (CDM) theory has serious issues on galactic and sub-galactic mass scales, even though it has been successful in interpreting a multitude of observational phenomena at cosmological scales.
These include the so-called missing satellite, core/cusp discrepancy, and too-big-to-fail (TBTF) problems.
One of the possible mechanisms to resolve these problems within the CDM paradigm is to rely on the effects of baryonic process such as the photoionization of neutral gas with  UV background and the impact of gas outflow mainly due to mainly supernova feedback on dark halo structures.

More radically, a possible solution that is well motivated from particle physics is to replace CDM with warm dark matter (WDM). 
The main difference between CDM and WDM is that while CDM particles have negligible thermal velocities, WDM ones have significant thermal velocities during the era of structure formation and thus these particles have a free-streaming length, giving rise to a cutoff matter power spectrum at larger wave numbers corresponding to small spatial scales.
The possible WDM candidates are the sterile neutrino~\citep[see, e.g.,][for a review]{Kusenko2009} or the gravitino~\citep[e.g.,][]{MMY1993}.
These candidates would have a mass in a few keV, leading to the suppression of structures on dwarf-galaxy mass scales.
Therefore, since the cutoff wavelength depends on the particle mass, investigating the nature of dark halo structures at low-mass scales leads to useful limits on the WDM particle mass, $m_{\rm WDM}$.

One of the powerful tools for constraining the WDM particle mass is Ly$\alpha$ forest observations~\citep[e.g.,][]{Narayanan2000,Viel2008,Viel2013}.
These observations allow us to study a matter power spectrum down to small scales and over a large redshift range.  
\citet{Viel2013} have observed the flux power spectrum of Ly$\alpha$ using the high-resolution spectra and have compared it to predicted ones from high-resolution $N$-body and hydrodynamical simulations, assuming that WDM masses are 1, 2, and 4 keV.
They concluded that in order to reproduce the observed power spectrum, WDM particles are required to have a lower limit of $m_{\rm WDM} \geq3.3$~keV ($2\sigma$ confidence level).

Dwarf spheroidal (dSph) galaxies also provide useful tests to distinguish between CDM and WDM or set constraints on particle mass of WDM. 
First, based on the difference of dark matter concentration between CDM and WDM subhalos, reproducing the maximum circular velocity of the Milky Way (MW) satellites can distinguish dark matter scenarios. This test is linked to the TBTF problem in CDM.
\citet{Lovetal2012}~performed $N$-body simulations with WDM and presented that the TBTF problem does not exist within the frameworks of WDM.
Also, \citet{PR2014}~showed that a WDM particle mass less than 4~keV can mitigate the TBTF problem in high-resolution simulations.
The second test is the method to resolve the missing satellite problem. 
Based on the difference of the abundance of subhalos within a host CDM and WDM dark halo, \citet{Lovetal2014} found a conservative lower limit of $m_{\rm WDM}>1.1$~keV in each $N$-body simulation.
However, these tests largely rely on the mass of the MW halo, where the estimates of the MW halo mass is still very uncertain with mass ranging from $\sim0.8\times10^{12} M_{\odot}$ to $\sim2\times10^{12} M_{\odot}$~\citep[e.g.,][]{Saketal2003, Deaetal2012, Boyetal2013, Kafetal2014, Pifetal2014}.  
For the case of the TBTF problem, CDM models can resolve the issue if the mass of the MW halo is less than $\sim1.5\times10^{12}M_{\odot}$~\citep{Wanetal2012}.
For the abundance of subhalos, if the MW halo mass is heavier than $1.8\times 10^{12} M_{\odot}$, only masses of WDM particles greater than 2~keV are accepted~\citep{Kenetal2014}.
Because of the dependence on the assumed MW halo mass and the uncertainty in it, the two tests above using the dwarf satellites are not yet sufficient to confine CDM or WDM scenarios.  

In this Letter, as yet another test on this issue from dwarf-galaxy scales, we revisit and consider the dark matter surface density, which appears nearly constant and independent of galaxy luminosity discovered by previous studies~\citep[e.g.,][]{Donato2009, Genetal2009, Salucci2012, KF2014,Burkert2015}. 
We propose that the re-definition and adoption of this surface density estimated within a radius of maximum circular velocity directly allow us to obtain the limits on particle masses of WDM.
This work is in part based on the kinematical analysis of the MW and Andromeda (M31) dSphs, for which more detailed descriptions will be presented elsewhere \citep*[][in preparation]{HC2015}.

\section{Dark matter surface density within a radius of maximum circular velocity}
Recently, analyzing the data of velocity dispersions, rotation curves of HI gas, and weak lensing for samples of dSph, dwarf irregular, spiral, and elliptical galaxies, \citet{Donato2009} and \citet{Genetal2009} fitted these data based on cored dark matter density profiles, such as the Burkert and pseudo-isothermal profiles, and found that the product of the central core density and core radius, $\rho_0r_0$, which is proportional to the central dark matter surface density, is constant for all these galaxies.

While these works offer us important suggestions on the universality of the dark halo structure, we note that this definition of $\rho_0r_0$ cannot be applied directly to a system with a cusped density profile. 
Both $\rho_0$ and $r_0$ in such a case may be defined in a specific way depending on an assumed density profile, so there exists some ambiguities when we compare surface densities between different systems and also compare with predictions from CDM or WDM.

To adequately compare dark matter distributions being independent of their density profiles, we introduce a mean surface density of a dark halo within a radius of maximum circular velocity,
\begin{eqnarray}
\Sigma_{V_{\rm max}} &=& \frac{M(r_{\rm max})}{\pi r^{2}_{\rm max}},
\label{SVmax}
\end{eqnarray}
where 
\begin{eqnarray}
M(r_{\rm max}) &=& \int^{r_{\rm max}}_{0}4\pi \rho_{\rm dm}(r^{\prime}) r^{\prime 2}dr^{\prime}, \nonumber
\label{Mmax}
\end{eqnarray}
$\rho_{\rm dm}(r)$ indicates any dark matter density profiles, and $r_{\rm max}$ is a radius at maximum circular velocity, $V_{\rm max}$, of assumed dark halo profiles, $\rho_{\rm dm}$.
The surface density defined in equation~(\ref{SVmax}) is exactly proportional to the central dark matter surface density, $\Sigma_{V_{\rm max}} \propto \rho_{\ast}r_{\ast}$, where $\rho_{\ast}$ and $r_{\ast}$ are the central density and scale length of arbitrary dark halo profiles, respectively.
We note that $r_{\rm max}$ is a better-defined scale than a core radius, $r_0$, because the latter is uncertain for a cuspy density profile: $r_0$ may be defined as a half-mass radius or some fraction of it in such a case, but it then suffers from the ambiguity of at what outer boundary the halo is tidally limited, e.g., the Navarro--Frenk-- White (NFW) profile having a diverging total mass.

\begin{figure}[t!]
\centering
\includegraphics[width=80mm]{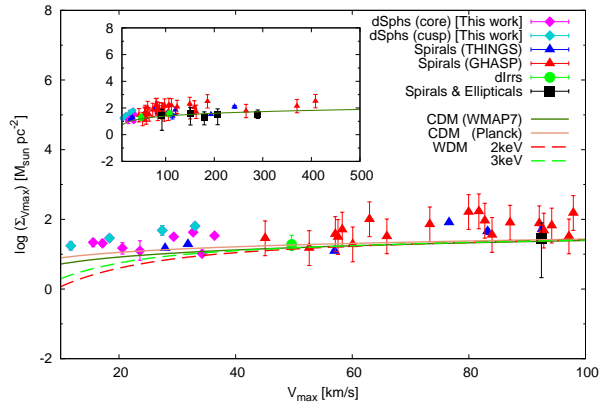}
\caption{Mean surface density of a dark halo, $\Sigma_{V_{\rm max}}$,  within a radius at maximum circular velocity, $V_{\rm max}$, as a function of $V_{\rm max}$ for different galaxies and Hubble types. 
While the inset shows all of the samples as well as CDM ({\it WMAP7}) prediction, the main panel highlights the plots only at $V_{\rm max}\leq100$~km~s$^{-1}$.  
Diamonds denote the MW and M31 dSphs derived in \citet*[][in preparation]{HC2015}, where magenta and cyan ones correspond to dSphs having a cored and cusped dark halo profile, respectively.
Blue and red triangles are the original \citet{Spaetal2008} sample of spiral galaxy data and nearby spirals in THINGS~\citep{deBetal2008}, respectively.
The data for dwarf irregular galaxies are indicated by green circles, and spirals and ellipticals investigated by weak lensing are labeled by black squares.
Solid lines show the results of CDM models with different cosmological parameters of {\it WMAP7} (dark green) and {\it Planck} (orange) measurements.
Red and green dashed lines denote WDM models with particle masses of 2 and 3~keV, respectively.}
\label{SVMAX2MAP} 
\end{figure}

Using this definition (\ref{SVmax}), we evaluate $\Sigma_{V_{\rm max}}$ within $r_{\rm max}$ taken from the literature based on the HI gas rotation curve of late- and early-type spirals with pseudo-isothermal dark halos~\citep{deBetal2008,Spaetal2008}, and dwarf irregulars with Burkert dark halos~\citep{Genetal2005,Genetal2007}, and the galaxy--galaxy weak lensing from spiral and elliptical galaxies with Burkert profiles (analyzed by \citealt{Donato2009}; data from \citealt{Hoeetal2005}).
For smaller, dwarf-galaxy scales, we analyze the stellar kinematics in the seven MW (Carina, Fornax, Sculptor, Sextans, Draco, Leo~I, and Leo~II) and five M31 (And~I, And~II, And~III, And~V, and And~VII) dSphs and construct axisymmetric mass models for these dark-matter-dominated dwarf galaxies~\citep*[][in preparation, which considers more generalized models than our previous work, \citealt{HC2012}]{HC2015}. 
For all samples, except for dSphs, we use the dark halo parameters (central density and core radius) estimated by the above original papers.

For dSphs, we employ the axisymmetric Jeans equations as in \citet{HC2012}, but we consider the effect of an additional velocity anisotropy of tracer stars, $\beta_z=1-\overline{v^2_z}/\overline{v^2_R}$, to determine the global shape of a dark halo, where $\overline{v^2_z}$ and $\overline{v^2_R}$ are velocity dispersions of stars in cylindrical coordinates (See \citealt*[][in preparation]{HC2015}, for more details). Together with $\beta_z$, we adopt an inner slope of the dark halo density profile as a free parameter as well as the parameterization of central density and scale length of a dark halo, apply our models to the two-dimensional line-of-sight velocity dispersion map for the above dSphs, and obtain the best-fit dark halo parameters.
We have found that whilst most of the dSphs have cored or shallower cusped dark halos, the Draco, Leo~I, And~III, and And~V dSphs show a steep inner density slope; not all of the dSphs have cored density profiles in their dark halos, as already suggested in \citet{HC2012}.

The symbols with error bars in Figure~\ref{SVMAX2MAP} show the estimated $\Sigma_{V_{\rm max}}$ of the above data sample as a function of $V_{\rm max}$.
It is found that even though dark halos in each galaxy sample are assumed to have different mass profiles (cusp or core) and are estimated by independent methods, the dark matter surface density within $r_{\rm max}$ is sufficiently {\it constant} across a wide range of galaxy masses. 

We discuss probable systematic uncertainties for the disk and dSph galaxies, respectively.
For the disk galaxies, we investigate the impact of the uncertainties in the adopted stellar mass-to-light ratio, $M_{\ast}/L_{\ast}$, on $\Sigma_{V_{\rm max}}$. 
\citet{deBetal2008} assumed the three $M_{\ast}/L_{\ast}$s and fitted dark halo parameters for each assumption. 
Adopting these parameters, we confirm that these uncertainties in $M_{\ast}/L_{\ast}$ do not largely affect the estimation of $\Sigma_{V_{\rm max}}$, varying it by a factor of three at maximum.  
For dSphs, we consider the influences of baryonic components and the velocity anisotropy of stars on the estimation of $\Sigma_{V_{\rm max}}$.
Since the fraction of the stellar mass in the total mass of the dSphs is so small, ranging from $10^{-1}$ to $10^{-3}$, that the estimated values of $\Sigma_{V_{\rm max}}$ may not at least change beyond a factor of 10, even if these small effects of baryonic components are taken into account.
In our models, we assume a {\it constant} velocity anisotropy of their member stars and find that this anisotropy is almost independent of other model parameters, such as a central density and scale length as well as an inner slope of a dark halo~\citep*[][in preparation]{HC2015}.
Therefore, the observational evidence for the constancy of $\Sigma_{V_{\rm max}}$ is robust.

\section{Comparison with dark matter models}
The use of the dark matter surface density defined by using $V_{\rm max}$ and $r_{\rm max}$ allows us to compare the results from the data sample with those from cosmological $N$-body simulations within the $\Lambda$-dominated CDM and WDM models.
We calculate $\Sigma_{V_{\rm max}}$ from the outcomes of the $N$-body simulations as follows.

First, we adopt the NFW dark matter density profile:
\begin{equation}
\rho(r) = \frac{\rho_s}{(r/r_s)(1+r/r_s)^2},
\label{NFW2}
\end{equation}
which can reproduce CDM~\citep[e.g.,][]{NFW1996,NFW1997} and WDM~\citep[e.g.,][]{Schneider2012,Lovetal2014} dark halos within a wide halo mass range.

Second, we calculate the scale density, $\rho_s$, and virial radius, $r_{\rm vir}$, for a NFW dark halo at the present time ($z=0$) from cosmological context:
\begin{eqnarray}
\rho_s &=& \frac{\rho_{\rm crit}\Delta_{m}}{3}\frac{c^3}{\ln(1+c)-c/(1+c)}, \\
r_{\rm vir} &=& \Bigl(\frac{3M_{\rm vir}}{4\pi\rho_{\rm crit}\Delta_m}\Bigr)^{1/3},
\end{eqnarray}
where $\Delta_{m}=97$ is a so-called virial overdensity at $z=0$~\citep{Praetal2012}, $c$ is a concentration parameter defined as $c=r_{\rm vir}/r_s$, $\rho_{\rm crit}=3H_0^2/8\pi G$ is a critical density of the current universe, and $H_0$ denotes the Hubble constant at $z=0$. 

\begin{figure}[t!]
\centering
\includegraphics[width=80mm]{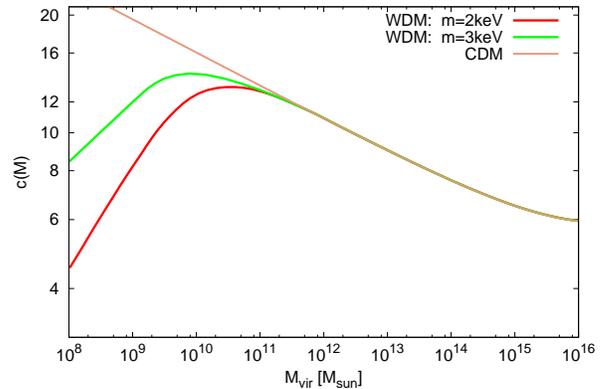}
\caption{Mass--concentration relations for dark halos of CDM (orange) and WDM with particle masses of 2~keV (red) and 3~keV (green).}
\label{CMvir} 
\end{figure}

Third, using the mass--concentration relation (MCR) for a dark halo, we obtain the scale radius, $r_s$.
For CDM models, we adopt two different MCRs at $z=0$ based on $N$-body simulations with slightly different cosmological parameters:
\begin{eqnarray}
c_{\rm W} = 10.5\Bigr(\frac{M_{\rm vir}}{10^{12}h^{-1}M_{\odot}}\Bigl)^{-0.08} \Bigr[1 + 0.15\Bigr(\frac{M_{\rm vir}}{10^{15}h^{-1}M_{\odot}}\Bigl)^{0.5}\Bigl]&&  \nonumber \\ 
{\rm (for\hspace{2mm}WMAP7)}&&,  \nonumber \\
c_{\rm P} = 10.7\Bigr(\frac{M_{\rm vir}}{10^{12}h^{-1}M_{\odot}}\Bigl)^{-0.11} \Bigr[1 + \Bigr(\frac{M_{\rm vir}}{2.4\times10^{16}h^{-1}M_{\odot}}\Bigl)^{0.4}\Bigl]&&\nonumber  \\ 
{\rm (for\hspace{2mm}Planck)}&&  \nonumber,
\end{eqnarray}
where $h$ is a dimensionless Hubble constant.
The MCR with cosmological parameters of {\it WMAP7} is derived by~\citet{Praetal2012}, whilst another one is found by~\citet{Klyetal2014}.
For WDM models, we adopt the method to derive MCR using the power spectra developed by \citet[hereafter S14]{Schneider2014}.
In order to evaluate the power spectrum of WDM models, the transfer function formula of the WDM universe by \citet{Viel2005} is adopted:
\begin{equation}
T_{\rm WDM}(k) = [1+(\alpha k)^{2\mu}]^{-5/\mu},
\end{equation}
with $\mu=1.12$ and
\begin{equation}
\alpha = 0.049\Bigr[\frac{m_{\rm WDM}}{{\rm keV}}\Bigl]^{-1.11}\Bigr[\frac{\Omega_{\rm WDM}}{0.25}\Bigl]^{0.11} \Bigr[\frac{h}{0.7}\Bigl]^{1.22} \hspace{3mm} h^{-1}{\rm Mpc},
\end{equation}
where $k$ is a wavenumber and $\Omega_{\rm WDM}$ denotes a cosmological density parameter for WDM.
Then, the {\it linear} power spectrum in the case of WDM is derived by
\begin{equation}
P_{\rm WDM}(k) =  T^2_{\rm WDM}(k)P_{\rm CDM}(k).
\end{equation}
Using the above power spectrum and MCR's fitting function described by equation (24) in \citet{Klyetal2014}, MCRs in the case of 2 and 3~keV particle masss of WDM are derived by comparing with the $N$-body simulations (see S14 for more details).
Figure~\ref{CMvir} displays the MCRs derived from CDM models with {\it Planck} data and WDM models with 2 and 3~keV particle masses at redshift 0. 
While CDM models predict that a concentration parameter monotonically increases with decreasing halo mass, WDM halos have a clear turnover at a low-mass halo.  

Finally, using the dark matter density profiles obtained by the above procedure, we compute $V_{\rm max}$ and $r_{\rm max}$, and then calculate $\Sigma_{V_{\rm max}}$ for CDM ({\it WMAP} and {\it Planck}) and WDM (particle masses of 2 and 3~keV) models.
The solid and dashed lines in Figure~\ref{SVMAX2MAP} are predicted $\Sigma_{V_{\rm max}}$ versus $V_{\rm max}$ for four cases of the dark matter models.
We emphasize that, surprisingly, these theoretical lines can well reproduce the data at high-mass scales, even though we do not perform any fitting to the data; actually, the data points (including low-mass scales) appear slightly above even the theoretical line giving the highest $\Sigma_{V_{\rm max}}$ (CDM with {\it Planck} parameters), which may suggest there are still some uncertainties in structural parameters for dark halos or even cosmology, but the following results for the limit of WDM remain unaffected.
For all mass ranges, CDM scenarios give a reasonable match to the data, where the results based on {\it WMAP} and {\it Planck} data are not so different at low-mass scales.
This is only due to the slight effect of the difference of cosmological parameters on the concentration of dark halos. In particular, $V_{\rm max}$ or $r_{\rm max}$ can depend on both the rms amplitude of linear mass fluctuation in a sphere of $8h^{-1}$~Mpc, $\sigma_8$, and the spectral index of the initial power spectrum, $n_s$ \citep{PR2014}.  
On the other hand, WDM ones are inconsistent with the universality of $\Sigma_{V_{\rm max}}$ at dwarf-galaxy mass scales.
This different feature between CDM and WDM models at the low-mass end stems from the properties of the halo concentrations at low halo-mass scales. 
From Figure~\ref{CMvir}, the concentrations of WDM halos differ substantially from those of CDM halos at low halo-mass scales, thereby the value of $\Sigma_{V_{\rm max}}$ decreases prominently at small $V_{\rm max}$ corresponding to dwarf galaxies.

\begin{figure}[t!]
\centering
\includegraphics[width=80mm]{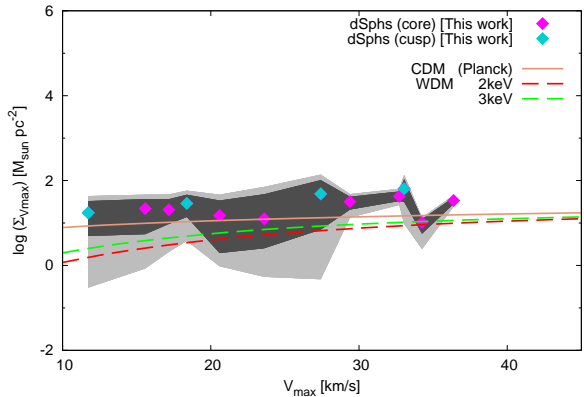}	
\caption{Same as Figure~\ref{SVMAX2MAP}, but for dSphs only. Black and gray shaded regions are $3\sigma$ and $5\sigma$ error areas, respectively, for each sample.}
\label{SVer} 
\end{figure}

We also investigate how far WDM models deviate from the data.
Figure~\ref{SVer} is the same as Figure~\ref{SVMAX2MAP}, but zooms in the dwarf-galaxy scale and plots $3\sigma$ and $5\sigma$ error areas of each sample. 
Focusing on $V_{\rm max} < 20$ km~s$^{-1}$ where the difference in $\Sigma_{V_{\rm max}}$ emerges significantly, even a 3~keV particle mass of WDM appears to deviate beyond a $3\sigma$ error in our sample.
Following this result, in order to be in agreement with the universality of $\Sigma_{V_{\rm max}}$ from the data, WDM particles need to be heavier than 3~keV.

\section{SUMMARY AND CONCLUSIONS}
In this Letter, we define and estimate a mean surface density of a dark halo within a radius of maximum circular velocity, $\Sigma_{V_{\rm max}}$ and apply it to dark halos in various galaxies.
This physical value reflects the central dark matter surface density, thus not strongly affected by the influence of external disturbances.
We find that even though dark halo parameters in each galaxy sample are derived based on the assumption of different dark halo density profiles and determined by several independent methods, this surface density is universal across a wide luminosity range of galaxies, irrespective of dark matter distribution in each of the galaxies.

Furthermore, this constancy is found to be useful for constraining dark matter models, especially the particle mass of WDM, because $\Sigma_{V_{\rm max}}$ is directly derivable from the theoretical predictions based on $N$-body simulations.
Our results show that at larger values of $V_{\rm max}$ corresponding to high halo-mass scales, this constancy for real galaxies can be successfully reproduced in both CDM and WDM models, even though we do not perform any fitting to the data. 
However, $\Sigma_{V_{\rm max}}$ derived from WDM largely deviates from the observational constancy at dwarf-galaxy mass scales, whilst that of CDM is in good agreement with it. 
This is because there is a prominent difference between MCR for CDM and WDM at low halo-mass scales.
In order to be in agreement with this constancy, we have found that WDM particles need to be heavier than 3~keV.

This Letter proposes an astrophysical test on the nature of dark matter.
The key dark halo mass range to distinguish dark matter models in this test is the lowest halo-mass scales, such as those of classical and ultra-faint dwarf galaxies. 
Wide-field spectroscopic studies of these galaxies over large areas with the planned Prime Focus Spectrograph to be mounted on the Subaru Telescope and high-precision spectroscopy for a plenty of much fainter stars with the Thirty Meter Telescope will provide a better determination of dark halo properties of the MW and M31 dSphs, and thus allow us to increase our knowledge about the basic properties of dark matter.

\acknowledgments
The authors thank the referee for her/his constructive comments that have helped us to improve our paper.
This work is supported in part by a Grants-in-Aid for Scientific Research from the Japan Society for the Promotion of Science (JSPS; No. 26-3302 for KH).
This work has been supported in part by a Grant-in-Aid for Scientific Research (25287062) of the Ministry of Education, Culture, Sports, Science and Technology in Japan and by the JSPS Core-to-Core Program ``International Research Network for Dark Energy.''



\begin{thebibliography}{}
\bibitem[Boylan-Kolchin et al.(2013)]{Boyetal2013} Boylan-Kolchin, M., Bullock, J.~S., Sohn, S.~T., Besla, G., \& van der Marel, R.~P.\ 2013, \apj, 768, 140 
\bibitem[Burkert(2015)]{Burkert2015} Burkert, A.\ 2015, arXiv:1501.06604 
\bibitem[de Blok et al.(2008)]{deBetal2008} de Blok, W.~J.~G., Walter, F., Brinks, E., et al.\ 2008, \aj, 136, 2648 
\bibitem[Deason et al.(2012)]{Deaetal2012} Deason, A.~J., Belokurov, V., Evans, N.~W., et al.\ 2012, \mnras, 425, 2840 
\bibitem[Donato et al.(2009)]{Donato2009} Donato, F., Gentile, G., Salucci, P., et al.\ 2009, \mnras, 397, 1169 
\bibitem[Gentile et al.(2005)]{Genetal2005} Gentile, G., Burkert, A., Salucci, P., Klein, U., \& Walter, F.\ 2005, ApJL, 634, L145 
\bibitem[Gentile et al.(2009)]{Genetal2009} Gentile, G., Famaey, B., Zhao, H., \& Salucci, P.\ 2009, \nat, 461, 627 
\bibitem[Gentile et al.(2007)]{Genetal2007} Gentile, G., Salucci, P., Klein, U., \& Granato, G.~L.\ 2007, \mnras, 375, 199 
\bibitem[Hayashi \& Chiba(2012)]{HC2012} Hayashi, K., \& Chiba, M.\ 2012, \apj, 755, 145 
\bibitem[K. Hayashi \& M. Chiba (2015)]{HC2015} Hayashi, K., \& Chiba, M. 2015, in preparation 
\bibitem[Hoekstra et al.(2005)]{Hoeetal2005} Hoekstra, H., Hsieh, B.~C., Yee, H.~K.~C., Lin, H., \& Gladders, M.~D.\ 2005, \apj, 635, 73 
\bibitem[Kafle et al.(2014)]{Kafetal2014} Kafle, P.~R., Sharma, S., Lewis, G.~F., \& Bland-Hawthorn, J.\ 2014, \apj, 794, 59 
\bibitem[Kennedy et al.(2014)]{Kenetal2014} Kennedy, R., Frenk, C., Cole, S., \& Benson, A.\ 2014, \mnras, 442, 2487 
\bibitem[Klypin et al.(2014)]{Klyetal2014} Klypin, A., Yepes, G., Gottlober, S., Prada, F., \& Hess, S.\ 2014, arXiv:1411.4001 
\bibitem[Kormendy \& Freeman(2014)]{KF2014} Kormendy, J., \& Freeman, K.~C.\ 2014, arXiv:1411.2170 
\bibitem[Kusenko(2009)]{Kusenko2009} Kusenko, A.\ 2009, PhR, 481, 1 
\bibitem[Lovell et al.(2014)]{Lovetal2014} Lovell, M.~R., Frenk, C.~S., Eke, V.~R., et al.\ 2014, \mnras, 439, 300 
\bibitem[Lovell et al.(2012)]{Lovetal2012} Lovell, M.~R., Eke, V., Frenk, C.~S., et al.\ 2012, \mnras, 420, 2318 
\bibitem[Moroi et al.(1993)]{MMY1993} Moroi, T., Murayama, H., \& Yamaguchi, M.\ 1993, PhLB, 303, 289 
\bibitem[Narayanan et al.(2000)]{Narayanan2000} Narayanan, V.~K., Spergel, D.~N., Dav{\'e}, R., \& Ma, C.-P.\ 2000, ApJL, 543, L103 
\bibitem[Navarro et al.(1996)]{NFW1996} Navarro, J.~F., Frenk, C.~S., \& White, S.~D.~M.\ 1996, \apj, 462, 563 
\bibitem[Navarro et al.(1997)]{NFW1997} Navarro, J.~F., Frenk, C.~S., \& White, S.~D.~M.\ 1997, \apj, 490, 493 
\bibitem[Piffl et al.(2014)]{Pifetal2014} Piffl, T., Scannapieco, C., Binney, J., et al.\ 2014, \aap, 562, AA91 
\bibitem[Polisensky \& Ricotti(2014)]{PR2014} Polisensky, E., \& Ricotti, M.\ 2014, \mnras, 437, 2922 
\bibitem[Prada et al.(2012)]{Praetal2012} Prada, F., Klypin, A.~A., Cuesta, A.~J., Betancort-Rijo, J.~E., \& Primack, J.\ 2012, \mnras, 423, 3018 
\bibitem[Sakamoto et al.(2003)]{Saketal2003} Sakamoto, T., Chiba, M., \& Beers, T.~C.\ 2003, \aap, 397, 899 
\bibitem[Salucci et al.(2012)]{Salucci2012} Salucci, P., Wilkinson, M.~I., Walker, M.~G., et al.\ 2012, \mnras, 420, 2034 
\bibitem[Schneider(2012)]{Schneider2012} Schneider, A., Smith, R.~E., Macci{\`o}, A.~V., \& Moore, B. 2012, \mnras, 424, 684
\bibitem[Schneider(2014)]{Schneider2014} Schneider, A.\ 2014, arXiv:1412.2133 
\bibitem[Spano et al.(2008)]{Spaetal2008} Spano, M., Marcelin, M., Amram, P., et al.\ 2008, \mnras, 383, 297 
\bibitem[Viel et al.(2013)]{Viel2013} Viel, M., Becker, G.~D., Bolton, J.~S., \& Haehnelt, M.~G.\ 2013, PhRvD, 88, 043502 
\bibitem[Viel et al.(2008)]{Viel2008} Viel, M., Becker, G.~D., Bolton, J.~S., et al.\ 2008, PhRvL, 100, 041304 
\bibitem[Viel et al.(2005)]{Viel2005} Viel, M., Lesgourgues, J., Haehnelt, M.~G., Matarrese, S., \& Riotto, A.\ 2005, PhRvD, 71, 063534 
\bibitem[Wang et al.(2012)]{Wanetal2012} Wang, J., Frenk, C.~S., Navarro, J.~F., Gao, L., \& Sawala, T.\ 2012, \mnras, 424, 2715 
\end{thebibliography}
\end{document}